\documentclass[twocolumn,showpacs,superscriptaddress,floatfix,aps,prl]{revtex4}
\usepackage{times}
\usepackage{amsmath}
\usepackage{graphicx}

\begin{document}
\date{\today}
\title{Complexity of 2D random laser modes at the transition \\ from weak scattering to Anderson localization}

\author{C. Vanneste}
\email[Contact: ]{vanneste@unice.fr}
\author{P. Sebbah}

\affiliation{Laboratoire de
Physique de la Mati{\`{e}}re Condens{\'{e}}e/CNRS UMR
6622/Universit{\'{e}} de Nice - Sophia Antipolis,
\\Parc Valrose, 06108, Nice Cedex 02, France}

\begin{abstract}
 The spatial extension and complexity of the eigenfunctions of an open finite-size two-dimensional (2D) random system are systematically studied for a random collection of systems ranging from weakly scattering to localized. The eigenfunctions are obtained by introducing gain in the medium and pumping just above threshold. All lasing modes are found to correspond to quasimodes of the passive system, for all regimes of propagation. We demonstrate the existence of multipeaked quasimodes or necklace states in 2D at the transition from localized to diffusive, resulting from the coupling of localized states.
\end{abstract}


\maketitle
Transport in random media is driven by the nature of the underlying eigenmodes. Propagation is diffusive when the modes extend spatially, while spectral level overlap occurs in transmission spectra. As the degree of the overlap decreases, transport is inhibited and modes become spatially localized \cite{Thouless}. The theory of Anderson localization predicts a transition between localized and extended eigenstates for spatial dimensions larger than two \cite{SPS}. Renewed interest in the localization transition has been boosted by the active ongoing search for localization of Bose-Einstein condensate in laser speckle fields \cite{Aspect,Inguscio,Skipetrov08}, the recent observations of the slowing down of diffusion in ultrasounds \cite{Page}, microwave \cite{Genack} and time-resolved optical \cite{Storzer} experiments, and new theoretical progresses \cite{Skipetrov06,Garcia} towards an analytical description of the metal insulator transition (MIT). The question of the spatial extent of the modes near the Anderson transition is also central in random lasers \cite{CaoWRM03,Lagendijk07}, where the threshold is still high. The threshold may vary by orders of magnitude between localized systems where the modes are spatially confined and diffusive systems where the modes are extended.

If the non-coexistence of extended and localized states was recently confirmed numerically \cite{Brndiar}, recent experimental observations of necklace states in nominally localized optical \cite{Wiersma05} and microwave \cite{Sebbah06} one-dimensional layered systems have attracted much attention. Modes overlapping both in space and frequency may couple and form multipeaked extended states even in the localized regime \cite{Bliokh}. Although scarce, they are predicted to play an outsized role in transport \cite{PendryPhysC} in contrast to isolated localized states \cite{Azbel}. Pendry argued that the picture of 1D necklace states should generalize to 2D and 3D localized random media \cite{PendryPhysC}. However, besides earlier calculations in a percolation model, no observations of necklace states in higher dimensions than one were reported \cite{PSheng91,Dasgupta}. In this context, the marginal dimension of two is of special interest since a pseudo-MIT is possible in 2D finite-sized opened samples where transport can be ballistic, diffusive (in contrast to 1D) or localized \cite{PRB}. Besides, the spatial profile of the modes is readily accessible as opposed to 3D, as shown in a recent observation of localized states within a 2D random microwave cavity \cite{Microwave}. The evidence of necklace states would therefore shine a new light on the Anderson transition, suggesting a possible scenario at the transition where coupling between increasing number of states could lead to a quantum percolation-type of transition \cite{Drabold}. This is to be compared with the filament-like fractal picture of the modes at the transition suggested by Aoki \cite{Aoki}, which lead to the extensive use of multifractal analysis to explore the MIT \cite{Janssen}.

Fundamental changes in the nature of the quasimodes are also expected in open random systems, as their spatial extension grows beyond the sample dimensions and their linewidth broadens with increasing leakage. This is analogous to quantum systems with an increasing degree of opening \cite{Stockman}, where the wavefunction becomes complex-valued with its standing wave component being progressively replaced by a component traveling toward the opened boundaries. The intensity distribution of the spatial pattern of the mode is expected to cross over from Porter-Thomas to Rayleigh distribution when new leakage channels open \cite{Shapiro96}. The complexity factor, $q^2=\langle Im(\Psi)^2\rangle/\langle Re(\Psi)^2\rangle$ \cite{Weaver}, together with the phase rigidity, $\rho=(1-q^2)/(1+q^2)$ \cite{Brouwer03}, was introduced in the field of quantum chaos to quantify the degree of complexity of the eigenfunction, $\Psi$, and the mutual influence of neighboring resonances \cite{Savin}. The complexity factor crosses over from 0 for real standing wavefunctions to 1 for purely traveling waves. To the best of our knowledge, it has never been used as a probe to characterize the transition from localized to extended states in finite disordered systems.

In this letter, we explore the nature of the quasimodes of 2D random media when scattering strength is increased. To obtain the quasimodes, the passive random system is embedded in a gain matrix and pumped just above the threshold. Single-mode lasing with resonant feedback is observed in any regime between extended and localized, with lasing modes corresponding systematically to a quasimode of the passive system. The spatial extension of the quasimodes, their complexity factor and the lasing threshold are calculated for a statistical ensemble of random configurations for each value of the scattering strength. These quantities reveal the change of regime and the transition from diffusive to localized. A detailed analysis of the phase probability distribution for each mode reveals multipeaked wavefunctions in the vicinity of the transition when the localization length is comparable to the sample size. These extended multipeaked quasimodes correspond to coupling of localized isolated states.

We consider a two-dimensional random collection of parallel dielectric cylinders with infinite extension, radius $r=60$ nm and refractive index $n$, embedded in a background matrix of index 1. Volume fraction is $\phi$ = 40 \% and system size is $L^2$ = 5x5 $\mu$m$^2$. Maxwell equations for TM polarization are modeled using the finite-difference time-domain method \cite{Taflove}.  Open boundary conditions are approximated by perfectly matched layer absorbing boundaries. The index of refraction $n$ is varied from 1.05 to 2.0, in step of 0.05, corresponding to scattering mean free path ranging from 50 $\mu$m to 0.1 $\mu$m. For most of this range, modes are short lived with strong spectral overlap, preventing individual excitation of a mode at its eigenfrequency by a monochromatic source in order to obtain its wavefunction. We showed recently that, when operating just above threshold, the first lasing mode of an active random system corresponds to a quasimode of the passive system, even in weakly scattering systems \cite{Vanneste07}. Introducing gain and pumping just above the threshold is therefore an alternative to select a quasimode of the passive cavity and obtain its spatial distribution, even in a system where modal overlap dominates. To model the gain, we couple the population equations of a four level atomic system to the Maxwell equations via the polarization equation \cite{Vanneste01}. The gain naturally selects the mode with the longest lifetime and the best spectral overlap with the gain curve.

\begin{figure}
\centering
\includegraphics[width=8cm,angle=0]{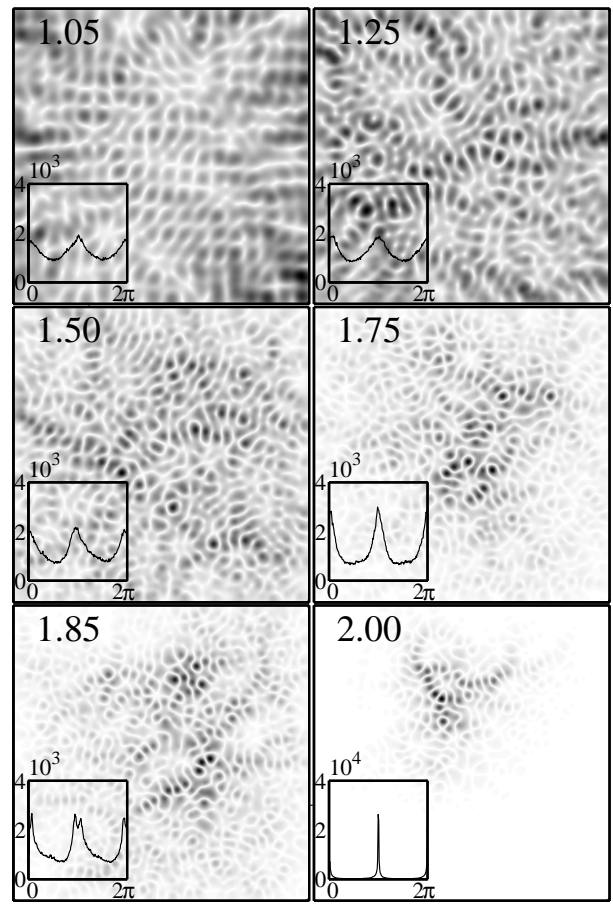}
\caption{Spatial distribution of the magnitude of the lasing mode just above the threshold, as the refractive index of the scatterers is increased from $n$=1.05 to 2.00 (random configurations are not necessarily identical). Each frame shows in inset the phase distribution between 0 and 2$\pi$ of the corresponding quasimodes. Note the double peaked distribution for $n$=1.85.} \label{Fig1}
\end{figure}

We study 150 random configurations, with an average of 10 per value of refractive index, $n$. For each configuration, the pump power is adjusted just above the threshold to observe single-mode lasing. Following the method using the spatial correlation introduced in \cite{Vanneste07}, we check that the lasing modes are the quasimodes of the passive system which provide the necessary feedback for lasing. Six spatial distributions of quasimodes corresponding to increasing values of the refractive index, $n$, are shown in Fig.~\ref{Fig1}, illustrating different degrees of spatial extension of the wavefunctions within the system. As modes extend toward the system boundaries with decreasing refractive index, they acquire an imaginary part corresponding to a traveling wave component, in contrast with localized modes, which are purely standing waves. This is seen in the phase probability distribution between 0 and $2\pi$. As shown in insets in Fig.~\ref{Fig1}, the phase probability distribution is peaked around 0 and $\pi$ when the mode is localized ($n$=2), while it is more uniformly distributed in the extended case. Note that for values of the refractive index as low as 1.05, scattering is weak and the field is rather concentrated at the edges of the system. In that case, residual reflection either at the boundaries or from the PML layers may not be negligible and result in periodic patterns, similar to those of a Fabry-Perrot cavity, as seen in the first frame of Fig.~\ref{Fig1}. We checked that above $n$=1.10, this effect is insignificant and lasing is solely due to multiple scattering within the system.

\begin{figure}
\centering
\includegraphics[width=8cm,angle=0]{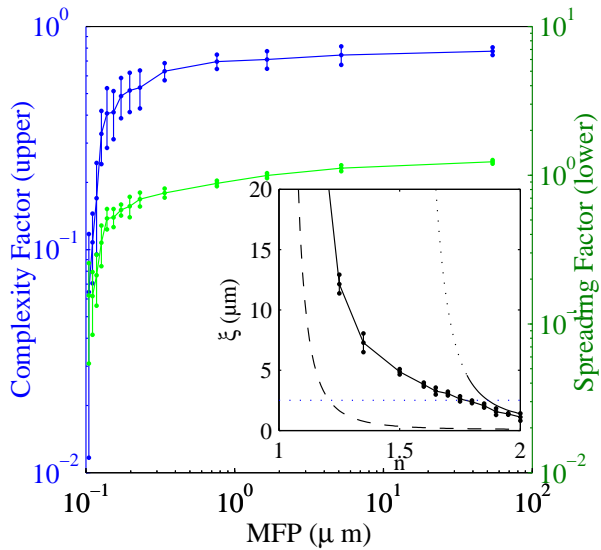}
\caption{Color online. (upper) Complexity factor, $q^2$ and (lower) spreading factor, $\eta$, averaged over sample configuration versus scattering mean free path, $\ell$. The fluctuations around the average, $\pm\langle{q^2-\langle{q^2}\rangle}\rangle^{1/2}$, and $\pm\langle{\eta^2-\langle{\eta}\rangle}\rangle^{1/2}$, are represented by the bars. Inset: Localization length $\xi$ versus index of refraction, $n$, calculated from the averaged lasing threshold (full line) and from independent scattering theory (full line where relevant ($\xi\leq L/2$), dotted otherwise). The dashed line represents the mean free path, $\ell$. The horizontal dotted line correspond to $\xi=L/2$}. \label{Fig2}
\end{figure}
The two main features observed as the scattering strength is reduced are the spatial expansion of the eigenfunctions and the increase of their imaginary parts associated with leakage at the open boundaries. To quantify these two characteristics, the quasimodes are described in terms of their spreading factor and their complexity factor. We define the spreading factor as $\eta=\int\int\tilde{A}(\vec{r})\tilde{A}(\vec{r'})|\vec{r}-\vec{r'}|^2d^2\vec{r}d^2\vec{r'}$, where the field amplitude $A(\vec{r})$ is normalized $\tilde{A}(\vec{r})=A(\vec{r})/[\int\int A^{2}(\vec{r})d^2\vec{r}]^{1/2}$.  The spreading factor is itself normalized to unity for uniform spatial distribution of the field. It measures the degree of spatial extension of the energy within the system alike the participation ratio for instance. However, due to the weighting factor $|\vec{r}-\vec{r'}|^2$, it also enables us to distinguish systems with spatial localization of energy inside the system, as in Fig.~\ref{Fig1} for $n$=2, from concentration of energy at the boundaries of the system, as in Fig.~\ref{Fig1} for $n$=1.05. Indeed, it can be less than 1 for spatially localized modes, or larger than one when energy is distributed near the system edges. This is reminiscent of distributed feedback lasers \cite{Kogelnick} where in the over-coupled regime (corresponding to $\eta<1$), energy is concentrated inside the laser as a result of strong feedback from scatterers, while in the under-coupled regime ($\eta>1$), energy is concentrated at the edges since the lasing modes result from scattering at the boundaries in order to maximize the gain volume \cite{Cao06}. The spreading factor and the complexity factor are computed for each mode \cite{PhaseShift} and averaged over sample configurations for each value of $n$. They are shown in Fig.~\ref{Fig2} as a function of the scattering mean free path, $\ell$, calculated using Mie theory for infinite cylinders of refractive index $n$. As the complexity factor explores values between 0.78 and 0.06, it shows clearly two different regimes, with a crossover around $\ell$=0.14 $\mu$m corresponding to $n$=1.8. A transition occurs around the same value of $\ell$ for the spreading factor, which ranges between 1.23 and 0.16. This value certainly is not universal and depends on the sample size but it should correspond to a localization length, $\xi$, of the order of the sample size. We confirm this hypothesis by calculating $\xi$ directly from the value of the lasing threshold averaged over sample realizations, $\langle P\rangle$. Indeed, the spectral width of the modes, $\Gamma$, resulting from leakage at the boundaries is given by $\Gamma=\Gamma_0\cdot\exp(-L/\xi)$ \cite{Microwave}. It is also directly proportional to the lasing threshold, $P$, since at threshold, losses are compensated by gain. Therefore $\xi=L/(\ln\langle{P_0}\rangle-\ln\langle{P}\rangle)$,where $\langle{P}\rangle$ designates the average over sample configurations for each value of $n$ and $\ln \langle{P_0}\rangle$=20.15 is obtained by extrapolating $\langle{P}\rangle$ at $n=1$. The dependence of $\xi$ on refractive index, $n$, is shown in the inset in Fig~\ref{Fig2} and is compared to the theoretical expression in the limit of independent scattering, as given in \cite{Microwave} by $\xi_{th}=\ell \exp[\pi Re(k_{eff})\ell/2]$, where $\ell$ is the mean free path and $k_{eff}$ is the effective wavenumber. Both curves join at the crossover, $n$=1.8, which corresponds to $\xi\sim L/2=$2.5 $\mu$m, where the two expressions for $\xi$ and $\xi_{th}$ start to be valid.

\begin{figure}
\centering
\includegraphics[width=8cm,angle=0]{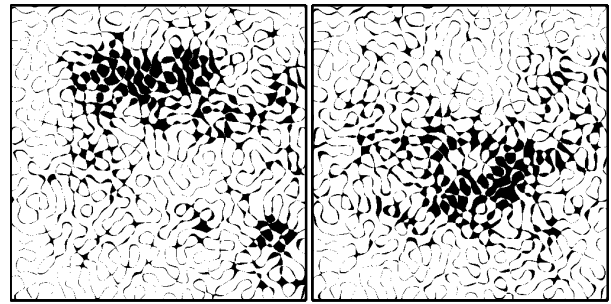}
\caption{Phase spatial distributions for phase values around (a) $0.94\pi$ and (b) $1.06\pi$.}
\label{Fig3}
\end{figure}

\begin{figure}
\centering
\includegraphics[width=8cm,angle=0]{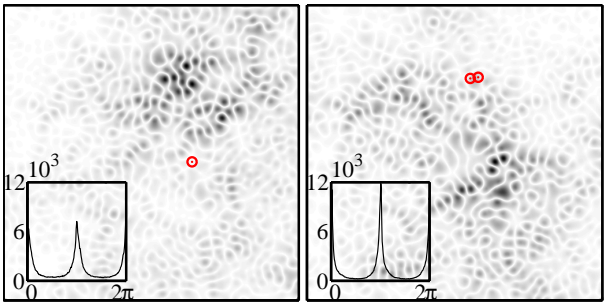}
\caption{Spatial distribution of the magnitude of the quasimode together with its phase distribution (in inset) for two different local perturbations of the random system of Fig.~\ref{Fig1} for $n$=1.85, when excited at the eigenfrequency of the original mode. The locations of the removed scatterers are shown by the circles.}
\label{Fig4}
\end{figure}

Also shown in Fig~\ref{Fig2} are the fluctuations of the complexity factor, $\pm\langle{q^2-\langle{q^2}\rangle}\rangle^{1/2}$, and the spreading factor, $\pm\langle{\eta^2-\langle{\eta}\rangle}\rangle^{1/2}$. A significant increase of these fluctuations is seen at the crossover. We find that these larger fluctuations are correlated with the occurrence of peculiar phase distributions around $n$=1.8, such as the one shown in the inset of Fig.~\ref{Fig1} for $n$=1.85. Two narrow peaks are distinctly seen in the phase distribution at $0.94\pi$ and $1.06\pi$. The spatial distributions of the phase for values comprised in a $\pi/10$ window around each of these two peaks are shown in Fig.~\ref{Fig3}a and b. These two distributions delimitate two distinct spatial regions associated with standing components of the mode, which oscillate at the eigenfrequency of the mode but with a phase lag, $\delta\phi=0.12\pi$. This phase lag suggests that this mode results from the coupling between two distinct modes localized on each of the regions of Fig.~\ref{Fig3}. This would be the analog of the symmetric or anti-symmetric solutions to the coupled oscillators problem in the presence of leakage, which introduces a phase lag different from $0$ or $\pi$ between the components of the hybridized mode. To identify each of the two components of the original double-peaked mode of Fig.~\ref{Fig1} ($n$=1.85), we use a perturbation approach where scatterers are removed in one of the two regions of Fig.~\ref{Fig3} to selectively separate the two contributions. Each perturbed system is excited at the resonant frequency of the original unperturbed mode. The corresponding field distributions are shown in Fig.~\ref{Fig4}a and b. They reproduce the local features of each peak of the mode of Fig.~\ref{Fig1}, but extend far beyond. The corresponding phase distributions are now single-peaked. Note that layers of randomly distributed scatterers (not shown) were added at the boundaries of the perturbed system in order to increase the lifetime of the mode of Fig.~\ref{Fig4}a, which would be impossible to excite otherwise due to strong leakage at the boundaries.

We find that hybridized quasimodes with two or three peaks arise almost systematically at $n$=1.8 and 1.85. The modes observed here demonstrate the existence of necklace states in 2D random media. They may play a crucial role at the transition from localized to diffusive. Two different mechanisms were proposed, which may coexist \cite{Drabold}. One possible mechanism is a gradual spatial expansion of the mode as scattering strength is diminished, with a progressive increase of the localization length. In a different scenario analogous to a percolation process, coupling of localized states could lead to extended structure as the spectral density of states increases to form necklace states \cite{PSheng06}. The multipeaked phase distributions and correspondingly multipeaked modes found here with high probability around the transition, suggest that  percolation is a possible scenario leading to a rapid crossover from localized to diffusive regimes in random media.

We thank D. Savin, O. Legrand and F. Mortessagne for fruitful discussions. This work was supported by the Centre National de la Recherche Scientifique (PICS $\#2531$ and PEPS07-20) and the Groupement de Recherches IMCODE.

{}


\begin{thebibliography}{}
\bibitem{PhaseShift} The phase is globally shifted for each mode to ensure the orthogonality of real and imaginary parts of its wavefunction, which is required in the definition of the complexity factor, $q^2$.
\end{thebibliography}
\end{document}